# p-ORDER ROUNDED INTEGER-VALUED AUTOREGRESSIVE (RINAR(p)) PROCESS

By **M. Kachour**

**IRMAR, University of Rennes 1, France.**

**M. Kachour**

IRMAR Institut de Recherche en Mathématiques de Rennes

UMR 6626/ Université Rennes 1

campus scientifique de Beaulieu

263 Avenue de Général Leclerc

35042 RENNES CEDEX

Email :

maher.kachour@univ-rennes1.fr

Tel :

00.33.(0)2.23.23.63.98

**Abstract**

An extension of the **RINAR(1)** process for modelling discrete-time dependent counting processes is considered. The model **RINAR(p)** investigated here is a direct and natural extension of the real **AR(p)** model. Compared to classical **INAR(p)** models based on the thinning operator, the new models have several advantages: simple innovation structure ; autoregressive coefficients with arbitrary signs ; possible negative values for time series ; possible negative values for the autocorrelation function. The conditions for the stationarity and ergodicity, of the **RINAR(p)** model, are given. For parameter estimation, we consider the least squares estimator and we prove its consistency under suitable identifiability condition. Simulation experiments as well as analysis of real data sets are carried out to assess the performance of the model.

**AMS classification code:** Primary $62M10$, secondary $62M20$.

# 1 Introduction.

The integer-valued autoregressive, **INAR(1)**, process is introduced by Al-osh & Alzaid [2] (1987) to model non-negative integer-valued phenomena that evolve in time. These models are based on the binomial thinning operator, denoted $\circ$, see [16].

The **INAR(1)** process is defined by

$$X_t = a_1 \circ X_{t-1} + \varepsilon_t, \quad \forall\, t \in \mathbb{N}, \tag{1}$$

where

$$a_1 \circ X_{t-1} = \sum_{k=1}^{X_{t-1}} \xi_{1k}.$$

Here, the so-called counting series $(\xi_{1k})$ are independent and identically distributed (i.i.d.) Bernoulli random variables with success probability $a_1 \in [0,1]$ and $(\varepsilon_t)$ is a sequence of i.i.d. non-negative integer-valued random variables and independent of the counting series. Thus, $a_1 \circ X_{t-1}$ is a binomial random variable with $a_1$ and $X_{t-1}$ as parameters, $a_1 \circ X_{t-1} \rightsquigarrow B(X_{t-1}, a_1)$.

The **INAR(p)** is an analogue of equation (1) with $p$ lags. An **INAR(p)** process is recursively defined by

$$X_t = a_1 \circ X_{t-1} + a_2 \circ X_{t-2} + \ldots + a_p \circ X_{t-p} + \varepsilon_t, \quad \forall\, t \in \mathbb{N}, \tag{2}$$

where, for $i = 1, \ldots, p$,

$$a_i \circ X_{t-i} = \sum_{k=1}^{X_{t-i}} \xi_{ik}.$$

Here $(\xi_{1k})$, $i \in \{1, \ldots, p\}$ and $k \in \mathbb{N} \setminus \{0\}$ are independent Bernoulli-distributed variables, where $\xi_{1k}$ has success probability $a_i \in [0,1]$ and $(\varepsilon_t)$ is a sequence of i.i.d. non-negative integer-valued random variables and independent of all the counting series.

The general **INAR(p)** processes where first introduced by Al-osh & Alzaid [3] (1990) but Du & Li [7] (1991) proposed a different specification. In the specification of Du & Li [7] (1991), denoted **INAR-DL**, the autocorrelation structure of an **INAR(p)** process is the same as that of an **AR(p)** process, whereas it corresponds to the one of an **ARMA(p,p-1)** process in the specification of Al-osh & Alzaid [3] (1990), denoted **INAR-AA**.

In particular, an **INAR(2)** process follows the equation

$$X_t = a_1 \circ X_{t-1} + a_2 \circ X_{t-2} + \varepsilon_t, \tag{3}$$

where

$$a_1 \circ X_{t-1} = \sum_{k=1}^{X_{t-1}} \xi_{1k} \rightsquigarrow B(X_{t-1}, a_1) \tag{4}$$

and

$$a_2 \circ X_{t-2} = \sum_{k=1}^{X_{t-2}} \xi_{2k} \rightsquigarrow B(X_{t-2}, a_2). \tag{5}$$

Note that each variable $X_u$ is thinned twice : $a_1 \circ X_u$ for $X_{u+1}$ and $a_2 \circ X_u$ for $X_{u+2}$.

We distinguish two different models according to the construction of these thinning operations:

The **INAR(2)-AA** specification introduced by Al-osh & Alzaid [3] (1990). Here, the counting series are chosen such that for each $X_u$, the vector $(a_1 \circ X_u, a_2 \circ X_u, X_u - a_1 \circ X_u - a_2 \circ X_u)$ follows a trinomial distribution with parameters $(X_{t-j}; a_1, a_2, 1 - a_1 - a_2)$. An important consequence arises from this choice of the counting series: a moving average structure is induced and the autocorrelation function of the process is similar to that of a **ARMA(2,1)** process.

Du & Li [7] (1991) propose a modified specification, denoted **INAR(2)-DL**. The counting series are chosen such that $\{a_1 \circ X_u, a_2 \circ X_u, u \in \mathbb{Z}\}$ are independent. The correlation properties of this process are identical to the **AR(2)** model.

Estimators of the parameters of **INAR(p)** are provided by several authors. For $p = 1$ and under the assumption that the sequence $(\varepsilon_t)$ has a Poisson distribution Franke & Seligmann [9] (1993) analysed maximum likelihood. Du & Li [7] (1991) derived the limit distribution of the OLS estimator of $a = (a_1, \ldots, a_p)$. Brännaäs and Hellström [5] (2001) considered GMM estimation, Silva and Oliveira [15] (2005) proposed a frequency domain-based estimator of $a$. Drost & al. [6] (2008) provided an efficient estimator of the parameters, and in particular, showed that the **INAR(p)** model has the Local Asymptotic Normality property.

Thus, the one-step ahead forecast, based on the conditional expectations, for the previous models is beset by the problem that forecast values obtained will be real rather than integer-valued in all but very rare cases. A mapping into the discrete support of the series is readily obtained by applying Gaussian brackets (integer part of), or by rounding to the nearest integer; the latter will be employed along this paper.

The class of **INAR(p)** models has several drawbacks. Their innovation structure is complex, depending not only on the noise process $(\varepsilon_t)$, but also on the counting variables. The autoregressive coefficients $a_j$'s are restricted to the interval $[0, 1]$. In the case of an **INAR(1)** model for example, this restriction excludes negative autocorrelations.

We introduce the following model

$$X_t = \langle \sum_{j=1}^{p} \alpha_j X_{t-j} + \lambda \rangle + \varepsilon_t, \quad t \in \mathbb{N}, \tag{6}$$

where $\langle \cdot \rangle$ represents the rounding operator to the nearest integer, $(\varepsilon_t)$ is a sequence of centered i.i.d. integer-valued random variables, defined on a probability space $(\Omega, \mathcal{A}, \mathbb{P})$, $\lambda$ and $(\alpha_j)$ are real parameters. We call this model **RINAR(p)** (for rounded integer-valued autoregression).

**RINAR(p)** has many advantages compared to the previous **INAR** models. Its innovation structure is simple, generated only by the noise $(\varepsilon_t)$. Its one-step ahead least squares predictor is given by

$$\hat{X}_{T+1} = \mathbb{E}\left(X_{T+1} \mid X_s, s \leq T\right) = \langle \sum_{j=1}^{p} \alpha_j X_{t-j} + \lambda \rangle, \tag{7}$$

which is integer-valued by construction. We shall also see that the **RINAR(p)** model can produce autocorrelation functions as rich as those of real **AR(p)**, including negatives autocorrelations. Moreover, by construction the **RINAR(p)** model can analyze a time series with negative values, a situation not covered by any **INAR** model. Lastly, it is worth mentioning that the $\langle \cdot \rangle$ is in fact the natural operation often used in the collection of integer-valued series.

In this paper, we study in details the **RINAR(p)** model. First in section 2, we give conditions ensuring the stationarity and the ergodicity of the model. Next in section 3, we introduce the least squares estimator for the estimation of the model parameters. This estimator is proved consistent under suitable conditions on the model. Because of the discontinuity of the rounding operator, particular care is needed for both the formulation of the model identifiability condition and the computation of the least squares estimator. A specific algorithm for the last problem is introduced in section 4. We then present a small simulation experiment in section 5 to assess the performance of the estimator. In section 6, we analyze a well-known time series with **RINAR(p)** models where classical integer-valued models are unsuccessful. Finally, section 7 collects the proofs of all theorical results.

## 1.1 Some notations

The following notations and properties will be used along this paper. First, let us define $\mathbb{N} = \{x \in \mathbb{Z} : x \geq 0\}$, $\mathbb{Z}_{>0} = \{x \in \mathbb{Z} : x > 0\}$ and $\mathbb{Z}^{-} = \{x \in \mathbb{Z} : x \leq 0\}$.

Now, we introduce several useful properties of the rounding operator $\langle \cdot \rangle$. Note that $\langle a \rangle$ is clearly defined anywhere, unless if $a = k + \frac{1}{2}$ where $k \in \mathbb{Z}$. By convention, we take $\langle k + \frac{1}{2} \rangle = k + 1, k \in \mathbb{N}$ and $\langle k - \frac{1}{2}\rangle = k - 1, k \in \mathbb{Z}^-$. Note that $a \to \langle a \rangle$ is an odd function.

Let $\{a\}$ be the fractional part of $a \in \mathbb{R}$, $\{a\} \in [0;1[$. Here, the fractional part of a negative number is a positive as $\{a\} = \{-a\} = \{|a|\}$, for example $\{-1.23\} = \{1.23\} = 0.23$.

Let $s$ be the sign function defined by $s(a) = 1$ if $a \geq 0$, and $s(a) = -1$ if $a < 0$.

Then, for all $a \in \mathbb{R}$, we have:

$$a = \langle a \rangle + s(a)\{a\} - s(a)\mathbb{1}_{\{a\} \geq \frac{1}{2}}. \tag{8}$$

Let $[a]$ be the integer part of $a \in \mathbb{R}$, for example $[2.8] = 2$ and $[-6.3] = -6$.

Then, for all $a \in \mathbb{R}$, we have:

$$a = [a] + s(a)\{a\}. \tag{9}$$

The following lemma will be useful and it is a direct consequence of equations (8) and (9).

**Lemma 1.** *Let $x \in \mathbb{R}$ and $a, b \geq 0$.*

1. $|\langle x \rangle - x| \leq \frac{1}{2}$.

2. $|\langle x \rangle| = \langle |x| \rangle \leq |x| + \frac{1}{2}$.

3. $\langle a + b \rangle = c + \langle \{a\} + \{b\} \rangle$, where $c = \langle a \rangle + \langle b \rangle - \mathbb{1}_{\{a\} \geq \frac{1}{2}} - \mathbb{1}_{\{b\} \geq \frac{1}{2}}$.

4. $\{a + b\} = \{\{a\} + \{b\}\}$.

5. $\langle a \rangle = [a] + \langle \{a\} \rangle$.

## 2 Ergodicity and stationarity of the RINAR(p) process.

The study of the **RINAR**(p) process can be carried out though the following vectorized process

$$Y_t = \begin{pmatrix} X_t \\ X_{t-1} \\ \vdots \\ X_{t-p+1} \end{pmatrix} = \begin{pmatrix} \langle \sum_{j=1}^p \alpha_j X_{t-j} + \lambda \rangle + \varepsilon_t \\ X_{t-1} \\ \vdots \\ X_{t-p+1} \end{pmatrix}. \tag{10}$$

The process $(Y_t)$ formes an homogeneous Markov chain with state space $E = \mathbb{Z}^p$ and transition probability function

$$\pi(x, y) = \mathbb{P}(\varepsilon_1 = y_1 - \langle \sum_{j=1}^p \alpha_j x_j + \lambda \rangle)\, \mathbb{1}_{y_2 = x_1, \ldots, y_p = x_{p-1}}, \ \forall\ x = (x_j), y = (y_j) \in E. \tag{11}$$

The following proposition gives the conditions which ensure the ergodicity and the stationarity of the **RINAR**(p) process. For $x = (x_1, \ldots, x_p) \in \mathbb{R}^p$, let $\|x\|_1 = |x_1| + \ldots + |x_p|$. For any measure $\mu$ and function $g$ on $E$, we set $\mu(g) = \int g(x)d\mu(x)$.

**Proposition 1.** *Suppose that:*

1. The Markov chain $(Y_t)$ is irreducible.

2. For some $k > 1$, $\mathbb{E}|\varepsilon_t|^k < +\infty$.

3. $\sum_{j=1}^{p} |\alpha_j| < 1$.

*Then*

1. The **RINAR(p)** process $(Y_t)$ has an unique invariant probability measure $\mu$ which has a moment of order $k$ (i.e. $\mu(\|.\|_1^k) < \infty$).

2. For all $y \in E$ and $f \in L^1(\mu)$ we have

$$\frac{1}{n} \sum_{k=1}^{n} f(Y_k) \longrightarrow \mu(f), \quad \mathbb{P}_y \text{ a.s.}$$

where $\mathbb{P}_y$ denotes the conditional probability $\mathbb{P}(. \mid Y_0 = y)$.

## 3 Estimation of parameters

Let $\theta = (\alpha_1, \ldots, \alpha_p, \lambda) \in \mathbb{R}^{p+1}$. In this section, it is assumed that $\theta$ belongs to a compact parameters space $\Theta$, subset of $]-1, 1[^p \times \mathbb{R}$. Let $x = (x_1, \ldots, x_p)' \in \mathbb{R}^p$. By convention, we round the vector $x$ coordinate-wisely, i.e. $\langle x \rangle = (\langle x_1 \rangle, \ldots, \langle x_p \rangle)$. We note

$$f(x; \theta) = f(x_1, \ldots, x_p; \theta) = \langle \sum_{j=1}^{p} \alpha_j x_j + \lambda \rangle.$$

Then, the **RINAR**(p) model can be written in the following form

$$Y_t = \begin{pmatrix} f(Y_{t-1}; \theta) \\ X_{t-1} \\ \vdots \\ X_{t-p+1} \end{pmatrix} + \begin{pmatrix} \varepsilon_t \\ 0 \\ \vdots \\ 0 \end{pmatrix} = \langle MY_{t-1} + \xi \rangle + \eta_t = F(Y_{t-1}; \theta) + \eta_t,$$

where

$$M = \begin{pmatrix} \alpha_1 \ldots \alpha_p \\ I_{p-1} \quad 0 \end{pmatrix}, \quad \xi = \begin{pmatrix} \lambda \\ 0 \\ \vdots \\ 0 \end{pmatrix} \quad \text{and} \quad \eta_t = \begin{pmatrix} \varepsilon_t \\ 0 \\ \vdots \\ 0 \end{pmatrix}.$$

Let $X_{-P+1}, \ldots, X_0, \ldots, X_n$ be observations from the **RINAR(p)** process. For the estimation of the parameter $\theta$, we consider the least squares estimator defined by

$$\hat{\theta}_n := \arg\min_{\theta \in \Theta} \varphi_n(\theta), \tag{12}$$

where

$$\varphi_n(\theta) = \frac{1}{n} \sum_{t=1}^{n} (X_t - f(Y_{t-1}; \theta))^2 = \frac{1}{n} \sum_{t=1}^{n} (\|Y_t - F(Y_{t-1}; \theta)\|_1)^2. \tag{13}$$

Some notations are necessary. To the norm $\|\cdot\|_1$ on $\mathbb{R}^p$, we associate a norm on $(\mathbb{R}^p)^2$ by setting $|z| := \|y_1\|_1 + \|y_2\|_1$ for $z := (y_1, y_2)$ in $(\mathbb{R}^p)^2$, where $y_1$ and $y_2$ are vectors in $\mathbb{R}^p$. The actual value of the parameter is denoted $\theta_0 = (\alpha_1^*, \ldots, \alpha_p^*, \lambda^*)$ and $\mathbb{P}_{\theta_0}$ stands for the probability distribution of the chain $(Y_t)$ under the actual model. Moreover, any convergence $\xrightarrow{a.s.}$ means an $a.s.$ convergence under $\mathbb{P}_{\theta_0, x}$, which hold independently of the initial state $x$. Let us make the following assumptions.

## Assumption [H]

1. Under $\mathbb{P}_{\theta_0}$, the Markov chain $(Y_t)$ is irreducible,

2. for some $k \geq 2$, $\mathbb{E}|\varepsilon_t|^k < +\infty$,

3. $\sum_{j=1}^p |\alpha_j^*| < 1$, where $\alpha_j^*$ is the actual value of $\alpha_j$,

4. the parametric space $\Theta$, is a compact subset of $]-1, 1[^p \times \mathbb{R}$.

Assume that [H] holds. Therefore, under $\mathbb{P}_{\theta_0}$ and from Proposition 1, $(Y_t)$ has an unique invariant measure $\mu_{\theta_0}$ such that $\mu_{\theta_0}(\|.\|_1^k) < \infty$ for some $k \geq 2$. Moreover, the double chain $(Z_t)$ with $Z_t = (Y_t, Y_{t-1})$ has similar properties. So, the chain $(Z_t)$ has also an unique invariant measure denoted $\sigma_{\theta_0}$. As $\mu_{\theta_0}(\|.\|_1^k) < \infty$, it follows that $\sigma_{\theta_0}(|.|^k) < \infty$ for the $|.|$ norm on $E^2$ defined above. Let the functions

$$g(z; \theta) = (\|y - F(x; \theta)\|)^2, \ z = (x, y) \in E^2, \ \theta \in \Theta, \tag{14}$$

and

$$K(\theta) = \sigma_{\theta_0} g(.; \theta), \ \theta \in \Theta. \tag{15}$$

The following proposition give us the limit of the least squares estimating function where the convergence holds uniformaly on $\Theta$.

**Proposition 2.** *Assume that [H] holds. Then, for any $\theta \in \Theta$, we have*

1. $\varphi_n(\theta) \xrightarrow{a.s.} K(\theta)$;

2. $K(\theta) - K(\theta_0) = \mu_{\theta_0}\left(f(.; \theta) - f(.; \theta_0)\right)^2$.

*Moreover,*

$$\sup_{\theta \in \Theta} |\varphi_n(\theta) - K(\theta)| \xrightarrow{a.s.} 0.$$

The proofs of this propositions and of all forthcoming results are postponed to section 7.

Next, we consider the consistency problem of the least squares estimator. As in the **RINAR(1)** case ( see [12]), the identifiability of **RINAR(p)** model has a non-standard behavior. Because of rounding operations, the model identifiability depends whether the autoregressive coefficients $\alpha_j^*$ are rational or not. For two parameter vectors $\theta = (\alpha_1, \ldots, \alpha_p, \lambda)$, $\theta' = (\alpha_1', \ldots, \alpha_p', \lambda')$ of $\Theta$, we define their distance to be

$$d(\theta, \theta') = \max\left\{|\alpha_j - \alpha_j'|, 1 \leq j \leq p, |\lambda - \lambda'|\right\}. \tag{16}$$

## 3.1 Strong consistency of the least squares estimator when at least one of $\alpha_j^*$ is irrational

The following proposition adresses precisely the question of identifiability of the parameters of **RINAR(p)** for the case where there exists at least $j \in \{1, \ldots, p\}$ such that $\alpha_j^* \in \mathbb{R} \backslash \mathbb{Q}$.

We recall, the following function

$$f(x;\theta) = \langle \sum_{j=1}^{p} \alpha_j x_j + \lambda \rangle, \ \forall \ x = (x_1, \ldots, x_p) \in E \text{ and } \theta = (\alpha_1, \ldots, \alpha_p, \lambda) \in \Theta.$$

**Proposition 3.** *Assume that [H] holds. If there exists at least one $j \in \{1, \ldots, p\}$ such that $\alpha_j^*$ is irrational. Then,*

$$f(x;\theta) = f(x;\theta_0), \ \forall \ x \in E \iff \theta = \theta_0.$$

**Proposition 4.** *Assume that [H] holds. If there exists at least one $j \in \{1, \ldots, p\}$ such that $\alpha_j^*$ is irrational. Then, for all (sufficiently small) $\varepsilon > 0$, we have*

$$\inf_{\theta \in \Theta_\varepsilon} |K(\theta) - K(\theta_0)| > 0,$$

*where $\Theta_\varepsilon = \{\theta : d(\theta, \theta_0) \geq \varepsilon\}$.*

**Theorem 1.** *Assume that*

1. *there exists at least one $j \in \{1, \ldots, p\}$ such that $\alpha_j^*$ is an irrational number;*
2. *under $\mathbb{P}_{\theta_0}$, the Markov chain $(Y_t)$ is irreducible;*
3. *for some $k \geq 2$, $\mathbb{E}|\varepsilon_t|^k < +\infty$;*
4. *$\sum_{j=1}^{p} |\alpha_j^*| < 1$, where $\alpha_j^*$ is the actual value of $\alpha_j$;*
5. *the parametric space $\Theta$, is a compact subset of $]-1,1[^p \times \mathbb{R}$.*

*Therefore, the least squares estimators are strongly consistent estimators of the actual values of the parameters, i.e.*

$$\hat{\theta}_n \to \theta_0, \quad \mathbb{P}_{\theta_0} - a.s.$$

## 3.2 Strong consistency of the least squares estimator when all the $\alpha_j^*$ are rationals

First, we recall the main result of the identifiability problem for **RINAR(1)** model [12], defined by

$$X_t = \langle \alpha X_{t-1} + \lambda \rangle + \varepsilon_t = f(X_{t-1}; \theta) + \varepsilon_t.$$

Let $\theta_0 = (\alpha_0, \lambda_0)$ be the actual value of the model.

If $\alpha_0 = \frac{m}{q}$, where $m \in \mathbb{Z}$, $q \in \mathbb{Z}_{>0}$, $m$ and $q$ are taken to be coprime, then

$$f(x; \theta) = f(x; \theta_0), \ \forall \ x \in E \iff \alpha = \alpha_0 \text{ and } \lambda \in I_0.$$

$I_0$ is a $\frac{1}{q}$-length interval or $\frac{1}{2q}$-length interval where $\lambda_0 \in I_0$.

The length of $I_0$ depends on the parity of $q$ and the position of $\{\lambda_0\}$ :

- If $q$ is even, then $I_0$ is a $\frac{1}{q}$-length interval.

- If $q$ is odd, then we distinguish two sub-cases :

    - If $\{\lambda_0\} \in \left[0; \frac{1}{2q}\right[ \cup \left[\frac{2q-1}{2q}; 1\right[$, then $I_0$ is a $\frac{1}{2q}$-length interval.
    - If $\{\lambda_0\} \in \left[\frac{1}{2q}; \frac{2q-1}{2q}\right[$, then $I_0$ is a $\frac{1}{q}$-length interval.

For the **RINAR(p)** model, we suppose that, for $j = 1, \ldots, p$ we have $\alpha_j^* = \frac{a_j}{b_j}$, where $a_j \in \mathbb{Z}$ and $b_j \in \mathbb{N}^*$,

$a_j$ and $b_j$ are taken to be coprime $(a_j \wedge b_j = 1)$. Let $\theta_0 = (\alpha_1^*, \ldots, \alpha_p^*, \lambda^*)$ be the actual value of the **RINAR(p)** model defined by (6).

Next, we will show that **RINAR(p)** is equivalent to a **RINAR(1)** model with a rational parameter denoted $\nu_0$.

Let $y = (x_1, \ldots, x_p)' \in E$. Thus,

$$\sum_{i=1}^{p} \alpha_i^* x_i = \frac{1}{\prod_{j=1}^{p} b_j} \left( \sum_{l=1}^{p} A_l x_l \right), \tag{17}$$

where

$$A_l = a_l \prod_{j=1, j \neq l}^{p} b_j. \tag{18}$$

From the Bézout theorem, we get

$$A_1 \mathbb{Z} + \cdots + A_p \mathbb{Z} = d\mathbb{Z}, \tag{19}$$

where $d = A_1 \wedge \cdots \wedge A_p$ is the P.G.C.D. of $A_1, \ldots, A_p$. Then, there exists $x \in \mathbb{Z}$ such that

$$\sum_{i=1}^{p} \alpha_i^* x_i = \nu_0 x, \tag{20}$$

with

$$\nu_0 = \frac{d}{\prod_{j=1}^{p} b_j} \in ]-1, 1[. \tag{21}$$

We note that the numerator and denumerator of $\nu_0$ are not necessary coprime.

Then, we rewrite $\nu_0$ with its irreductible fraction form

$$\nu_0 = \frac{a}{b}, \tag{22}$$

where $a \in \mathbb{Z}$, $b \in \mathbb{N}^*$ and $a \wedge b = 1$.

**Proposition 5.** *Assume that [H] holds. Let $\theta = (\alpha_1^*, \ldots, \alpha_p^*, \lambda) \in \Theta$. It follows*

$$\forall\ y \in E,\ f(y;\theta) = f(y;\theta_0) \iff \forall\ x \in \mathbb{Z},\ \langle \nu_0 x + \lambda \rangle = \langle \nu_0 x + \lambda^* \rangle.$$

**Proposition 6.** *Assume that [H] holds. If for all $j = 1, \ldots, p$ we have $\alpha_j^* = \frac{a_j}{b_j}$, where $a_j \in \mathbb{Z}$ and $b_j \in \mathbb{N}^*$, then*

$$f(x;\theta) = f(x;\theta_0),\ \forall\ x \in E \iff \alpha_j = \alpha_j^*,\ \forall\ j = 1, \ldots, p\ \text{and}\ \lambda \in I_0.$$

$I_0$ *is a $\frac{1}{b}$-length interval or $\frac{1}{2b}$-length interval where $\lambda^* \in I_0$.*

The length of $I_0$ depends on the parity of $b$ and the position of $\{\lambda^*\}$ :

- If $b$ is even, then $I_0$ is a $\frac{1}{b}$-length interval.

- If $b$ is odd, then we distinguish two sub-cases :

    - If $\{\lambda^*\} \in \left[0; \frac{1}{2b}\right[ \cup \left[\frac{2b-1}{2b}; 1\right[$, then $I_0$ is a $\frac{1}{2b}$-length interval.
    - If $\{\lambda^*\} \in \left[\frac{1}{2b}; \frac{2b-1}{2b}\right[$, then $I_0$ is a $\frac{1}{b}$-length interval.

Next, in the order to simplify the presentation, we give full details only in the case where $b$ is even.

Let $\alpha_0 = (\alpha_1^*, \ldots, \alpha_p^*)$. From the above discussions, there exists $k_0 \in \{0, 1, \ldots, b-1\}$ such that

$$\{\lambda^*\} \in i_0 = \left[\frac{k_0}{b}, \frac{k_0+1}{b}\right[.$$

We define,

$$I_0 = \{\lambda : \langle \lambda \rangle = \langle \lambda^* \rangle\ \text{and}\ \{\lambda\} \in i_0\}\ \text{and}\ E_0 = \{\alpha_0\} \times I_0. \tag{23}$$

For example, we take the **RINAR(4)** model with

$$\alpha_*^1 = \frac{3}{25},\quad \alpha_*^2 = \frac{3}{8},\quad \alpha_*^3 = \frac{1}{5}\quad \alpha_*^4 = -\frac{1}{4}\quad \text{and}\ \lambda^* = 2.5.$$

It is follows, from (18), that

$$\prod_{j=1}^{4} b_j = 800,\quad A_1 = 480,\quad A_2 = 1500,\quad A_3 = 800\quad \text{and}\ A_4 = -1000.$$

So, we have $d = A_1 \wedge A_2 \wedge A_3 \wedge A_4 = 20$. Then, we get, from (21), that $\nu_0 = \frac{20}{800}$ and its irreductible fraction form

$$\nu_0 = \frac{1}{40} = 0.025.$$

Therefore, $b = 40$ is even, we have

$$E_0 = \left\{ \left(\frac{3}{25}, \frac{3}{8}, \frac{1}{5}, -\frac{1}{4}\right) \right\} \times [2.5, 2.525[.$$

**Proposition 7.** *Assume that [H] holds. If $\alpha_j^* = \frac{a_j}{b_j}$, $a_j \in \mathbb{Z}$ and $b_j \in \mathbb{N}^*$, $a_j$ and $b_j$ are coprime $\forall\ j = 1, \ldots, p$ and $b$ the denumerator of (22) is even, then for all (sufficiently small) $\varepsilon > 0$, we have*

$$\inf_{\theta \in \Theta_\varepsilon^0} |K(\theta) - K(\theta_0)| > 0,$$

*where $\Theta_\varepsilon^0 = \{\theta : d(\theta, E_0) \geq \varepsilon\}$.*

**Theorem 2.** *Assume that*

1. *$\alpha_j^* = \frac{a_j}{b_j}$, $a_j \in \mathbb{Z}$ and $b_j \in \mathbb{N}^*$, $a_j$ and $b_j$ are coprime $\forall\ j = 1, \ldots, p$,*

2. *$b$ the denumerator of (22) is even.*

3. *under $\mathbb{P}_{\theta_0}$, the Markov chain $(Y_t)$ is irreducible,*

4. *for some $k \geq 2$, $\mathbb{E}|\varepsilon_t|^k < +\infty$,*

5. *$\sum_{j=1}^p |\alpha_j^*| < 1$, where $\alpha_j^*$ is the actual value of $\alpha_j$,*

6. *the parametric space $\Theta$, is a compact subset of $]-1, 1[^p \times \mathbb{R}$.*

*Then we get $d(\hat{\theta}_n, E_0) \to 0$, $\mathbb{P}_{\theta_0}$ – a.s. In other words, $\hat{\alpha}_n$ is strongly consistent while $\hat{\lambda}_n$ converges to an interval of length $\frac{1}{b}$ containing $\lambda^*$.*

In the case where $b$ is odd, Theorem 2 still holds where $I_0$ (therefore $E_0$) is remplaced by the corresponding intervals as mentioned in Proposition 6.

We know that, from [12], for **RINAR(1)** model with $\alpha_0 = \frac{a_0}{b_0}$ such that $a_0$ and $b_0$ are coprime, the length of interval $I_0$ is equal to $\frac{1}{b_0}$. There is a natural question :

can we reduce, with $p > 1$ parameters, the length of the interval $I_0$ ?

Let us consider the **RINAR(2)** model with

$$\alpha_1^* = \frac{a_1}{b_1} \quad \text{and} \quad \alpha_2^* = \frac{a_2}{b_2}.$$

Then, we get

$$\nu_0 = \frac{d}{b_1 b_2} = \frac{a}{b} \quad \text{where} \quad d = a_1 b_2 \wedge a_2 b_1 \quad \text{and} \quad a \wedge b = 1.$$

Let $A$ be an interval in $\mathbb{R}$, we note by $|A|$ the length of $A$. We have $|I_0| = \frac{1}{b}$.

Thus

$$b/a\,(b_1 b_2) \quad \text{and} \quad a \wedge b = 1 \quad \text{then} \quad b/b_1 b_2.$$

It follows that

$$\frac{1}{b_1 b_2} \leq |I_l|.$$

The **RINAR(1)** model with the parameter $\alpha_0 = \alpha_1^*$ (resp. $\alpha_2^*$) produces the interval noted $A_1$ (resp. $A_2$). Then, $|A_1| = \frac{1}{b_1}$ and $|A_2| = \frac{1}{b_2}$. Let $B = A_1 \cap A_2$. We have

$$I_0 \subseteq B \implies |I_0| \leq |B| \leq \min\left\{\frac{1}{b_1}, \frac{1}{b_2}\right\}.$$

Finally,
$$\frac{1}{b_1 b_2} \leq |I_0| \leq \min\left\{\frac{1}{b_1}, \frac{1}{b_2}\right\}.$$

Therefore, by increasing the autoregression order $p$, we can actually expect a reduction of the interval $I_0$ where the parameter $\lambda$ remains unidentifiable.

# 4 A numerical method to calculate $\hat{\theta}_n$

To calculate the least squares estimator, we generalize the minimization algorithm proposed in [12]. As in **RINAR(1)** model, the initialization step is given by the Yule-Walker's method. Therefore, the generalized algorithm continues through successive dichotomous search steps.

First, we will explain, in a general way, the algorithm for $\varphi_n$ with a scalar $\theta$.
The aim is to find
$$\hat{\theta}_n = \arg \min_{\theta \in [a,b]} \varphi_n(\theta).$$

Here, the initial search interval is $[a, b]$ (i.e. $left = a$ and $right = b$). The objective function $\varphi_n$ is evaluated for every step of the search at three different points: $\hat{\theta}_k$ and their middel left ($mid-left$) and middel right ($mid-right$) points. According to the minimum value of $\varphi_n(\hat{\theta}_k)$, $\varphi_n(mid-left)$ and $\varphi_n(mid-right)$, $\hat{\theta}_k$, $left$ and $right$ change their actual value (see Figure 1). For example, if $\varphi_n(\alpha_k)$ is the minimum then $left$ takes the value of the $mid-left$ and $right$ takes the value of the $mid-right$. This process stops when $range = |rigth - left| \leq 0.001$.

The following pseudo-code defines the used dichotomous search of $\theta$.

```
left <- a; right <- b; range <- abs ( b - a );
mid-left <- ( left + previous_theta ) / 2;
mid-right <- ( right + previous_theta ) / 2;
while range > 0.001 do
 begin
   V_1 <- \Phi_n ( \theta );
   V_2 <- \Phi_n ( mid-left );
   V_3 <- \Phi_n ( mid-right );
   i <- j such that V_j is min (V_1, V_2, V_3)
   case i of
     1 : begin left <- mid-left; right <- mid-right end;
     2 : begin right <- \theta; \theta <- mid-left end;
     3 : begin left <- \theta; \theta <- mid-right end;
  range <- abs ( right - left )
 end;
```

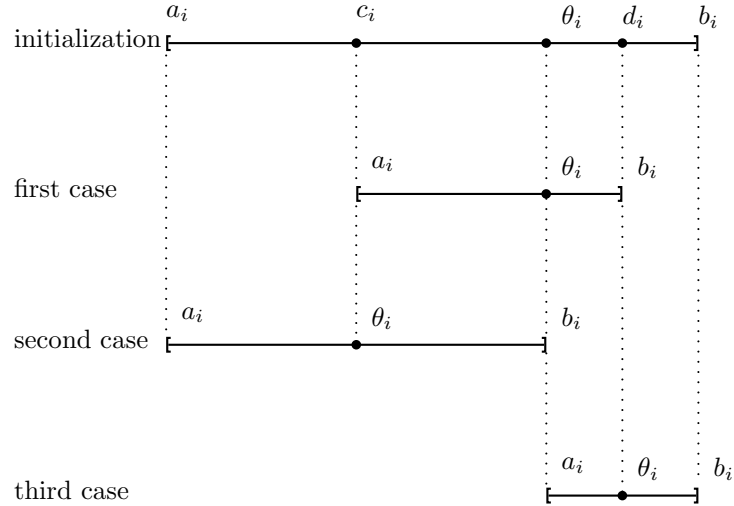

Figure 1: Dichotomous search for a scalar parameter $\theta$.

## 4.1 Initialization

As for **RINAR(1)**, we propose, the initial value $\hat{\theta}_0 = (\hat{\alpha}_{1,0}, \ldots, \hat{\alpha}_{p,0}, \hat{\lambda}_0)$ is defined as the Yule-Walker estimator as in a real **AR(p)** model :

$$\begin{pmatrix} \hat{\alpha}_{1,0} \\ \vdots \\ \hat{\alpha}_{p,0} \end{pmatrix} = \hat{R}_p^{-1} \hat{\rho}_p \quad \text{and} \quad \hat{\lambda}_0 = \bar{X}_n (1 - \sum_{j=1}^{p} \hat{\alpha}_{j,0}),$$

where $\hat{R}_p = [\hat{\rho}(i-j)]_{i,j=1}^{p}$ is the sample auto-correlation matrix, $\hat{\rho}_p = (\hat{\rho}(1), \ldots, \hat{\rho}(p))'$ and $\bar{X}_n$ the sample mean.

## 4.2 Successive dichotomy search steps

Now, the transition from $(\hat{\alpha}_{1,k}, \ldots, \hat{\alpha}_{p,k}, \hat{\lambda}_k)$ to $(\hat{\alpha}_{1,k+1}, \ldots, \hat{\alpha}_{p,k+1}, \hat{\lambda}_{k+1})$, for $k = 0, 1, 2, \ldots$
is done in $p+1$ phases. The first $p$ phases represent the passage of $\hat{\alpha}_{j,k}$ to $\hat{\alpha}_{j,k+1}$, $j = 1, \ldots, p$, where the initial search interval is $]-1, 1[$ (i.e. $left = -1$ and $right = 1$). The last one represent the passage of $\hat{\lambda}_k$ to $\hat{\lambda}_{k+1}$ where the intial search interval is defined to be $\hat{\lambda}_0 \pm 5|\hat{\lambda}_0|$, which seems large enough in most of situations. In every phase, we use the same algorithm described above, keeping whenever the results obtained in the phase before. This is the end of $(k+1)^{th}$ iteration.

The search stops when the results from two consecutive iterations are very close. More precisely, we stop at the $k^{th}$ iteration if :

$$d(\hat{\theta}_k, \hat{\theta}_{k+1}) = \max\left\{|\hat{\alpha}_{1,k} - \hat{\alpha}_{1,k+1}|, \ldots, |\hat{\alpha}_{p,k} - \hat{\alpha}_{p,k+1}|, |\hat{\lambda}_k - \hat{\lambda}_{k+1}|\right\} \leq 0.001. \qquad (24)$$

This stopping criterion is satisfied after few iterations. At the end of the iterations, we get the vector $\hat{\theta}_n = (\hat{\alpha}_n^1, \ldots, \hat{\alpha}_n^p, \hat{\lambda}_n)$ that minimizes our objective function $\varphi_n(\alpha_1, \ldots, \alpha_p, \lambda)$.

## 5   A simulation study

For this simulation study, we consider a **RINAR(4)** model. The error variable, say $\varepsilon_1$, is generated as $\varepsilon_1 = Z_1 - Z_2$ where $Z_1$ and $Z_2$ are two i.i.d. Poisson random variables. We simulate, $n = 500$ observations, of **RINAR(4)** model with $\theta_0 = (\frac{3}{25}, \frac{3}{8}, \frac{1}{5}, -\frac{1}{4}, \frac{5}{2})$, values already used in section 3.2. After 500 independent replications, the average and the standard deviation of the sequence of the estimates $\hat{\theta}_n$ obtained, are computed and given in the following table.

| $\theta_0$ | mean | s.e. |
|---|---|---|
| $\alpha_*^1 = 0.12$ | 0.1204 | 0.0473 |
| $\alpha_*^2 = 0.375$ | 0.3687 | 0.0439 |
| $\alpha_*^3 = 0.2$ | 0.1932 | 0.0425 |
| $\alpha_*^4 = -0.25$ | $-0.2472$ | 0.0454 |
| $\lambda^* = 2.5$ | 2.548 | 0.2766 |

The histograms of these estimates are displayed in Figure 2. Note that inevitably, the $\alpha_j^*$'s used for simulations are rational. Therefore, from Theorem 2, the consistency of the least squares estimator $\hat{\theta}_n$ has to be judged within some inevitable fluctuations of $\hat{\lambda}_n$.

Thus, recall that for $\alpha_0 = (\alpha_*^1, \alpha_*^2, \alpha_*^3, \alpha_*^4) = (\frac{3}{25}, \frac{3}{8}, \frac{1}{5}, -\frac{1}{4})$ we get $\nu_0 = \frac{1}{40}$ (see Section 3.2).

We know that $\hat{\theta}_n \longrightarrow E_0 = \{\alpha_0\} \times I_0$ where $I_0$ is an interval around $\lambda^*$ of length $\frac{1}{40}$.

## 6   Analysis of the Fürth data

In this section, we consider the 505 counts of pedestrians on a city block observed every 5 seconds originally published by Fürth [10], see Figure 3. This data set is well known in the branching process literature, see Mills and Seneta [14] (1989). The counts vary from 0 to 7. The mean of the series is 1.59 and its variance 1.51. Figure 4 shows the sample ACF and PACF of the data. Note that in particular, the second partial autocorrelation is significantly negative.

### 6.1   The INAR(2) model of Jung and Termayne.

For the Fürth data, Jung and Termayne [11] fitted a **INAR(2)** model

$$X_t = a_1 \circ X_{t-1} + a_2 \circ X_{t-2} + b + \varepsilon_t,$$

where $a_1, a_2 \in [0, 1]$. Both the two **INAR(2)** specifications introduced in section 1 are considered. With the **INAR(2)-AA** specification, they have $\hat{a}_1 = 0.664$, $\hat{a}_2 = -0.119$ and $\hat{b} = 0.723$. The corresponding estimates for the **INAR(2)-DL** specification are : $0.808$, $-0.214$ and $0.646$. None of the specifications used yields a satisfactory model, because the fitted value $\hat{a}_2$ is negative which is not allowed in an **INAR(2)** model.

The authors [11] also fitted a **INMA(2)** model. The **INMA** class, as the **INAR** class, is based on the binomial thinnig operator. Its structure is

$$X_t = b_1 \circ \varepsilon_{t-1} + b_2 \circ \varepsilon_{t-2} + b + \varepsilon_t, \text{ where } b_1, b_2 \in [0, 1].$$

The **INMA** process has been studied by McKenzie [13] (1988). Al-osh & Alzaid [1] (1988) have proposed other specification of the McKenzie's model. For a comparaison of the two specification, we refer Brännäs and Hall [4] (2001). Jung & Termayne have used the McKenzie specification. Then, the Yule-Walker estimates for the parameters are $\hat{b}_1 = 1.008$, $\hat{b}_2 = 0.961$ and $\hat{b} = 0.536$. A main difficulty here is that the estimate for the first thinning parameter $b_1$ is outside the admissible parameter space $[0, 1]$.

## 6.2 Fit of a RINAR(2) model.

Here, by using the software R, we will try to fit a real autoregressive model for the Fürth data. Figure 4 shows that the first and the second sample autocorrelations, are more significant than the others and there exists a cutt-off after lag 2 in the partial autocorrelations. First, we consider a **AR(2)** model

$$X_t = a_1 X_{t-1} + a_2 X_{t-2} + b + \varepsilon_t.$$

Then, the Yule-Walker estimates for the parameters are $0.808$ for $a_1$ ($s.e.0.0434$), $-0.214$ for $a_2$ ($s.e.0.0435$) and $0.646$ for $b$ (the same estimates for the **INAR(2)-DL**). Both $a_1$ and $a_2$ are significant and the AIC value equals $1328.99$.

Now, we consider a **AR(3)** model

$$X_t = a_1 X_{t-1} + a_2 X_{t-2} + a_3 X_{t-3} + b + \varepsilon_t.$$

The Yule-Walker estimates for the parameters are $0.828$ for $a_1$ ($s.e.0.0442$), $-0.292$ for $a_2$ ($s.e.0.0561$), $0.0986$ for $a_3$ ($s.e.0.0443$) and $0.578$ for $b$. We note, $a_3$ is not very significant. As the associated AIC value $1326.05$ is very close to the previous one, we will consider the order $p = 2$ hereafter.

Therefore, we propose a **RINAR(2)** model

$$X_t = \langle \alpha_1 X_{t-1} + \alpha_2 X_{t-2} + \lambda \rangle + \varepsilon_t.$$

In order examine forecast results for this model, we reserved the 400 initial observations as a learning set (i.e., to estimate the parameters) and the 105 latest observations as a test set for forecasting. By the algorithm of section 4, we find $\hat{\theta}_n = (\hat{\alpha}_{1,n}, \hat{\alpha}_{2,n}, \hat{\lambda}_n) = (0.818, -0.23, 0.697)$.
The one-step least squares ahead forecast $\hat{X}_{T+1}$ of $X_{T+1}$ equals here :

$$\hat{X}_{T+1} = \langle \hat{\alpha}_1 X_T + \hat{\alpha}_2 X_{T-1} + \hat{\lambda} \rangle. \tag{25}$$

The forecast errors $\hat{\varepsilon}_{T+1} = X_{T+1} - \hat{X}_{T+1}$, $400 \leq T \leq 504$ are displayed on Figure 5.
The mean absolute error equals

$$MAE = \frac{1}{105} \sum_{T=400}^{504} |\hat{\varepsilon}_{T+1}| = 0.743. \tag{26}$$

This $MAE$ value is not fully satisfactory indicating that the other **RINAR(p)** models could be considered. Yet, the use of a **RINAR(2)** model turns to be more natural than the previously prposed **INAR(2)** or **INMA(2)** fit. In particular, the estimated model is consistent with regard to the domain of definition of the parameters which was not the case for the previous **INAR(2)** or **INMA(2)** models.

# 7 Proofs

**Proof of Proposition 1**

First, we define on $\mathbb{R}^p$ the functions $\varphi$ and $V$ with

$$x \mapsto \varphi(x) = \begin{pmatrix} |x_1| \\ \vdots \\ |x_p| \end{pmatrix}, \qquad x \mapsto V(x) = (\sum_{j=1}^p |x_j|)^k = \|x\|_1^k.$$

Since $V$ is positive and $\lim_{\|x\|_1 \to \infty} V(x) = \infty$, $V$ is therefore a Lyapunov function ( see Duflo [8]).
Let "$\leq$" be a partial order relation over $\mathbb{R}^p$ defined by

$$x \leq y \iff x_j \leq y_j, \quad \forall j = 1, \ldots, p,$$

for $x = (x_j)$ and $y = (y_j)$. We have

$$|\langle \sum_{j=1}^p \alpha_j X_{t-j} + \lambda \rangle + \varepsilon_t| \leq \sum_{j=1}^p |\alpha_j| |X_{t-j}| + |\lambda| + |\varepsilon_t| + \frac{1}{2},$$

then

$$\varphi(Y_t) \leq A\varphi(Y_{t-1}) + \zeta_t,$$

where

$$A = \begin{pmatrix} |\alpha_1| \ldots |\alpha_p| \\ I_{p-1} \quad 0 \end{pmatrix} \quad \text{and} \quad \zeta_t = \begin{pmatrix} |\lambda| + |\varepsilon_t| + \frac{1}{2} \\ 0 \\ \vdots \\ 0 \end{pmatrix},$$

with $I_{p-1}$ is the identity matrix of size $p-1$. Note that $\|\zeta_t\|_1 = |\lambda| + |\varepsilon_t| + \frac{1}{2}$.
Iterating this estimate, we get

$$\varphi(Y_{t+n}) \leq A^n \varphi(Y_t) + A^{n-1}\zeta_t + \ldots + A\zeta_t + \zeta_t,$$

as $\|\cdot\|_1$ is an increasing function over $(\mathbb{R}^p, \leq)$,

$$\|\varphi(Y_{t+n})\|_1 \leq \|A^n \varphi(Y_t) + A^{n-1}\zeta_t + \ldots + A\zeta_t + \zeta_t\|_1$$
$$\leq \|A^n \varphi(Y_t)\|_1 + \|A^{n-1}\zeta_t\|_1 + \ldots + \|A\zeta_t\|_1 + \|\zeta_t\|_1$$
$$\leq \|A^n \varphi(Y_t)\|_1 + S_n(A)\|\zeta_t\|_1,$$

where

$$S_n(A) = ||| A^{n-1} |||_1 + \ldots + ||| A |||_1 + 1,$$

with $||| \cdot |||_1$ is the operator norm associate to $\|\cdot\|_1$. As a result

$$V(Y_{t+n}) \leq \left( \|A^n \varphi(Y_t)\|_1 + S_n(A)(|\lambda| + |\varepsilon_t| + \frac{1}{2}) \right)^k.$$

Then

$$((\pi^n V)(x))^{\frac{1}{k}} \leq \left( \mathbb{E}_x \left( \|A^n \varphi(Y_t)\|_1 + S_n(A)(|\lambda| + |\varepsilon_t| + \frac{1}{2}) \right)^k \right)^{\frac{1}{k}}$$
$$\leq \left( \mathbb{E}_x \|A^n \varphi(Y_1)\|_1^k \right)^{\frac{1}{k}} + S_n(A)b$$
$$= \|A^n \varphi(x)\|_1 + S_n(A)b$$
$$\leq ||| A^n |||_1 \|x\|_1 + S_n(A)b,$$

where $b = \left(\mathbb{E}(|\lambda| + |\varepsilon_t| + \frac{1}{2})^k\right)^{\frac{1}{k}}$. We have $\mathbb{E}|\varepsilon_t|^k < \infty$ then $0 < b < \infty$.

We note that for any matricial norm we have $||| A^n |||^{\frac{1}{n}} \longrightarrow \rho(A)$ as $n \to \infty$, where $\rho(A)$ is the spectral radius of $A$. As $\sum_{j=1}^p |\alpha_j| < 1$, then $\rho(A) < 1$. It follows, there exists $n_0$ such that $\forall\, n \geq n_0$ we have $||| A^{n_0} ||| = \alpha' < 1$. So, from Cauchy's criterion we get $\sum ||| A^n |||_1$ is convergent. Therefore, we have

$$((\pi^n V)(x))^{\frac{1}{k}} \leq \alpha' \|x\|_1 + \beta = \|x\|_1 \left( \alpha' + \frac{\beta}{\|x\|_1} \right).$$

So, we get

$$\frac{(\pi^n V)(x)}{V(x)} \leq \left( \alpha' + \frac{\beta}{\|x\|_1} \right)^k$$

where $\alpha' < 1$ and $0 < \beta < \infty$. Therefore, $(Y_t)$ verifies the Lyapunov criterion, with $V(x) = \|x\|_1^k$ as Lyapunov function (see Duflo [8], proposition 2.1.6) and we have that $(Y_t)$ is irreducible. It follows that $(Y_t)$ is positive recurrent with an unique probability invariant measure $\mu$ with $\mu V < \infty$. The conclusion (2) follows the classical ergodic theorem for Markov chains. $\square$

Some recalls are necessary. $E = \mathbb{Z}^p$ is the state space of the Markov chain $(Y_t)$ defined by (10). For all $x = (x_1, \ldots, x_p) \in E$, let $\|x\|_1 = \sum_{j=1}^p |x_j|$. We define $|\cdot|$ the norme on $E^2$ by

$$\forall\, z = (x,y) \in E^2,\ |z| = \|x\|_1 + \|y\|_1.$$

As the assumption [H] holds, $\mu_{\theta_0}$ and $\sigma_{\theta_0}$ ( the respective unique invariant probability of the chain $(Y_t)$ and the double chain $Z_t = (Y_{t-1}, Y_t)$ under the actual model) have both a moment of order $k \geq 2$. Here we recall the following functions:

$$f(x; \theta) = \langle \sum_{j=1}^{p} \alpha_j x_j + \lambda \rangle, \ \forall \ x = (x_1, \ldots, x_p) \in E \text{ and } \theta = (\alpha_1, \ldots, \alpha_p, \lambda) \in \Theta,$$

$$F(x; \theta) = (f(x; \theta), x_1, \ldots, x_{p-1})',$$

$$g(z; \theta) = (\|y - F(x; \theta)\|_1)^2, \ \forall \ z = (x, y) \in E^2,$$

$$\text{and } K(\theta) = \sigma_{\theta_0} g(.; \theta).$$

In all of the following Proofs, we denoted a generic constant $c$ whose exact value can change during the mathematical development.

The following Lemma will be useful for the proof of Proposition 2.

**Lemma 2.** *For all $\theta \in \Theta$, we have*

1. $(\|F(.; \theta)\|_1)^2 \in L^1(\mu_{\theta_0})$.

2. $|g(.; \theta)| \in L^1(\sigma_{\theta_0})$

## Proof of Lemma 2

Note that, for all $a \in \mathbb{R}$, $|\langle a \rangle| \leq |a| + \frac{1}{2}$. Thus, by using the compactness of $\Theta$, for all $x = (x_1, \ldots, x_p)' \in E$, we get

$$\|F(x; \theta)\|_1 = |f(x; \theta)| + |x_1| + \ldots + |x_{p-1}|$$
$$\leq (|\alpha_1| + 1)|x_1| + \ldots + (|\alpha_{p-1}| + 1)|x_{p-1}| + |\alpha_p||x_p| + |\lambda| + \frac{1}{2}$$
$$\leq c \ (1 + \|x\|_1).$$

Then,

$$(\|F(x; \theta)\|_1)^2 \leq c \ (1 + \|x\|_1^2).$$

So, because $k \geq 2$ and $\mu_{\theta_0}(\|\cdot\|_1^k) < \infty$, the first conclusion follow.

Now, for all $z = (x, y) \in E^2$, we have

$$\|y - F(x; \theta)\|_1 \leq \|y\|_1 + \|F(x; \theta)\|_1$$
$$\leq c \ (1 + \|x\|_1 + \|y\|_1)$$
$$= c \ (1 + |z|),$$

it follows

$$g(z; \theta) \leq c \ (1 + |z|^2). \tag{27}$$

So, because $k \geq 2$ and $\sigma_{\theta_0}(|\cdot|^k) < \infty$, the second conclusion follow. $\square$

**Proof of Proposition 2**

1. First, we identify the limit of the least squares estimating function. Note that the contrast function $\varphi_n$, defined by (13), equals:

$$\varphi_n(\theta) = \frac{1}{n}\sum_{t=1}^{n} g(Z_t; \theta). \tag{28}$$

From Lemma 2, we have $g(.;\theta) \in L^1(\sigma_{\theta_0})$. Therefore, by the ergodic theorem for the double Markov chain $(Z_t)$, we get

$$\varphi_n(\theta) \xrightarrow{a.s.} \sigma_{\theta_0} g(.;\theta) = K(\theta).$$

2. Our aim is to prove that the function $K(\theta)$ satisfies

$$K(\theta) - K(\theta_0) = \mu_{\theta_0}\left(f(.;\theta) - f(.;\theta_0)\right)^2.$$

For this, we will show that $\varphi_n(\theta) - \varphi_n(\theta_0) \xrightarrow{a.s.} \mu_{\theta_0}\left(f(.;\theta) - f(.;\theta_0)\right)^2$.

The following definitions and notations will be used in the remainder of the proof.

Let $\mathcal{F} = (\mathcal{F}_n)_{n\geq 0}$ be the natural filtration associated to the **RINAR(p)** process where $\mathcal{F}_n = \sigma(\varepsilon_t, 0 \leq t \leq n)$ for $n \geq 1$, and $\mathcal{F}_0$ is the degenerated $\sigma$-algebra. If $(M_n)$ is a square integrable martingale w.r.t. $\mathcal{F}$, we denote by $([M]_n)$ its increasing process defined by:

$$[M]_0 = 0, \quad [M]_n = [M]_{n-1} + \mathbb{E}(\| [M]_n - [M]_{n-1} \|_1^2 \mid \mathcal{F}_{n-1}) \quad \text{for } n \geq 1.$$

Let $[M]_\infty = \lim[M]_n$. On $\{[M]_\infty < \infty\}$, we have $M_n \xrightarrow{a.s.} M_\infty$, where $M_\infty$ is a finite random variable. On $\{[M]_\infty = \infty\}$, we have $M_n/[M]_n \xrightarrow{a.s.} 0$ (See Duflo [8], Theorem 1.3.15, p. 20).

Recall that, under the actual model $\theta_0$, $Y_t = F(Y_{t-1}; \theta_0) + \eta_t$.

We denote $\Delta F_{t-1} = F(Y_{t-1}; \theta_0) - F(Y_{t-1}; \theta)$. It follows

$$\varphi_n(\theta) - \varphi_n(\theta_0) = \frac{A_n}{n} + \frac{B_n}{n}, \tag{29}$$

with:

$$A_n = \sum_{t=1}^{n} \|\Delta F_{t-1}\|_1^2 \quad \text{and} \quad B_n = 2\sum_{t=1}^{n}(\eta_t \mid \Delta F_{t-1}),$$

where $(\cdot \mid \cdot)$ is the scalar product associate to $\|\cdot\|_1$. We have that

$$\Delta F_{t-1} = \begin{pmatrix} f(Y_{t-1};\theta) - f(Y_{t-1};\theta_0) \\ 0 \\ \vdots \\ 0 \end{pmatrix}, \quad \text{so } \|\Delta F_{t-1}\|_1^2 = (f(Y_{t-1};\theta) - f(Y_{t-1};\theta_0))^2.$$

It follows that $M_n := \frac{B_n}{2} = \sum_{t=1}^{n} \varepsilon_t(f(Y_{t-1};\theta) - f(Y_{t-1};\theta_0))$. From Lemma 2, we have $|f(.;\theta)|^2 \in L^1(\mu_{\theta_0})$. Then, by the ergodic theorem, we get

$$\frac{A_n}{n} \xrightarrow{a.s.} \mu_{\theta_0}\left(f(.;\theta) - f(.;\theta_0)\right)^2. \tag{30}$$

It is simple to verifies that $M_n$ is a square integrable martinale. Its increasing process $[M]_n$ is equal to
$$[M]_n = \sum_{t=1}^{n} (f(Y_{t-1};\theta) - f(Y_{t-1};\theta_0))^2 \, \mathbb{E}|\varepsilon_t|^2.$$

As almost surely,
$$\frac{1}{n}[M]_n \longrightarrow \mu_{\theta_0}\left((f(.;\theta) - f(.;\theta_0))^2 \Gamma\right) \geq 0, \quad \text{where} \quad \Gamma = \mathbb{E}|\varepsilon_t|^2 < \infty,$$

it follows that $\frac{1}{n}M_n \longrightarrow 0$, therefore
$$\frac{B_n}{n} \longrightarrow 0. \tag{31}$$

3. Our aim is to prove that
$$\sup_{\theta \in \Theta} |\varphi_n(\theta) - K(\theta)| \xrightarrow{a.s.} 0.$$

Some notations are necessary. For any $z = (x,y) \in E^2$, let $|z| = \|x\| + \|y\|$.

Let $\mathbb{P}_n$ be empirical measure generated by the observations $Z_1, \ldots, Z_n$
$$\mathbb{P}_n(z) = \frac{1}{n}\sum_{i=1}^{n} \mathbb{1}_{Z_i = z}, \quad z = (x,y) \in E^2.$$

It follows, from equation (28), the contrast function $\rho_n$ equals:
$$\rho_n(\theta) = \mathbb{P}_n g(.;\theta). \tag{32}$$

From equation (27), we get
$$|g(z;\theta)| \leq A(z), \ \forall \ z = (x,y) \in E^2 \text{ and } \theta \in \Theta, \tag{33}$$

where
$$A(z) = c(1 + |z|^2), \quad \text{where c is a constant.} \tag{34}$$

Note that, $A \in L^1(\sigma_{\theta_0})$. Let $q > 0$ be fixed. It arises,
$$\begin{aligned}
|\varphi_n(\theta) - K(\theta)| &= |(\mathbb{P}_n - \sigma_{\theta_0})g| \\
&= |(\mathbb{P}_n - \sigma_{\theta_0})g\mathbb{1}_{|z|<q} + (\mathbb{P}_n - \sigma_{\theta_0})g\mathbb{1}_{|z|>q}| \\
&\leq |(\mathbb{P}_n - \sigma_{\theta_0})g\mathbb{1}_{|z|<q}| + \mathbb{P}_n(|g|\mathbb{1}_{|z|>q}) + \sigma_{\theta_0}(|g|\mathbb{1}_{|z|>q}) \\
&\leq |(\mathbb{P}_n - \sigma_{\theta_0})g\mathbb{1}_{|z|<q}| + \mathbb{P}_n(A(z)\mathbb{1}_{|z|>q}) + \sigma_{\theta_0}(A(z)\mathbb{1}_{|z|>q}).
\end{aligned}$$

We denote $p_z = \sigma_{\theta_0}(z)$ and $p_z^n = \mathbb{P}_n(z)$. Moreover,
$$\begin{aligned}
|(\mathbb{P}_n - \sigma_{\theta_0})g\mathbb{1}_{|z|<q}| &= |\sum_{|z|<q} g(z;\theta)(p_z^n - p_z)| \\
&\leq \sum_{|z|<q} A(z)|p_z^n - p_z|.
\end{aligned}$$

This means that
$$\sup_{\theta \in \Theta} |\varphi_n(\theta) - K(\theta)| = \sup_{\theta \in \Theta} |(\mathbb{P}_n - \sigma_{\theta_0})g| \leq \sum_{|z|<q} A(z)|p_z^n - p_z| + (\mathbb{P}_n + \sigma_{\theta_0})(A(z)\mathbb{1}_{|z|>q}).$$

When $n \to \infty$, we have $p_z^n \to p_z$ almost surely, then the finite sum $\sum_{|z|<q} A(z)|p_z^n - p_z| \to 0$, $a.s.$ Therefore,

$$\limsup_{n \to \infty} \sup_{\theta \in \Theta} |(\mathbb{P}_n - \sigma_{\theta_0})g| \leq 0 + 2\sigma_{\theta_0}(A(z)\mathbb{1}_{|z|>q}), \ a.s.$$

By making $q \nearrow \infty$, we get almost surely,

$$\limsup_{n \to \infty} \sup_{\theta \in \Theta} |(\mathbb{P}_n - \sigma_{\theta_0})g| = 0. \quad \square$$

Recall that $\Theta$ is compact. Thus, for all $\theta = (\alpha_1, \ldots, \alpha_p, \lambda) \in \Theta$, there exists two positive constants $A$ and $B$ such that $|\lambda| \leq B$ and $|\alpha_j| \leq A$, $\forall j = 1, \ldots, p$. $\{\cdot\}$ represents the fractional part operator. For all $a \in \mathbb{R}$, $\{a\} \in [0; 1[$ and $\{a\} = \{|a|\}$.

## Proof of Proposition 3

Our aim is to prove that if there exists at least one $j \in \{1, 2, \ldots, p\}$ such that $\alpha_j^* \in \mathbb{R} \setminus \mathbb{Q}$, then for all $\theta = (\alpha_1, \ldots, \alpha_p, \lambda) \in \Theta$, we have

$$\langle \sum_{i=1}^p \alpha_i x_i + \lambda \rangle = \langle \sum_{i=1}^p \alpha_i^* x_i + \lambda^* \rangle, \ \forall \ x = (x_i) \in E \implies \lambda = \lambda^* \text{ and } \alpha_i = \alpha_i^*, \ \forall i = 1, \ldots, p.$$

The idea is to prove that if $\theta \neq \theta_0$ then there exists $x_0 \in E$ such that $f(x_0; \theta) \neq f(x_0; \theta_0)$.

We assume, without loss of generality, $\alpha_1 \neq \alpha_1^*$. Let $x = (x_1, 0, \ldots, 0) \in E$. Thus, by using the compactness of $\Theta$, we get

$$|f(x; \theta) - f(x; \theta_0)| \geq |\alpha_1 - \alpha_1^*||x_1| - 2B - 1.$$

Therefore, if $(\alpha_i) \neq (\alpha_i^*)$, then there exists $x_0 \in E$ such that $|f(x_0; \theta) - f(x_0; \theta_0)| > 0$, this is a contradiction. Moreover, if $x_0 = 0_E$, then $f(x_0; \theta) = f(x_0; \theta_0)$ implies $\langle \lambda \rangle = \langle \lambda^* \rangle$.

In order to simplify the notations, we assume that $\alpha_1^* \in \mathbb{R} \setminus \mathbb{Q}$. Let $y = (x, 0, \ldots, 0) \in E$.

We assume, without loss of generality, $\alpha_1^* > 0$, $\lambda \geq \lambda^* \geq 0$ and $\{\lambda\}, \{\lambda^*\} \in \left[0; \frac{1}{2}\right[$.

It follows, $\lambda - \lambda^* = \{\lambda\} - \{\lambda^*\}$ and

$$f(y; \theta) = f(y; \theta_0) \Rightarrow \langle \{\alpha_1^* x\} + \{\lambda\} \rangle = \langle \{\alpha_1^* x\} + \{\lambda^*\} \rangle. \tag{35}$$

As $\alpha_1^* \in \mathbb{R} \setminus \mathbb{Q}$, it arises $(\{\alpha_1^* x\})_{x \in \mathbb{Z}}$ is dense in $[0; 1[$. If $\{\lambda\} \neq \{\lambda^*\}$, then there exists $y_0 \in \mathbb{Z}$ such that

$$\{\alpha_1^* y_0\} + \{\lambda\} \geq \frac{1}{2} \text{ and } \{\alpha_1^* y_0\} + \{\lambda^*\} < \frac{1}{2}.$$

Finally, there exists $x_0 = (y_0, 0, \ldots, 0) \in E$ such that $f(x_0; \theta) \neq f(x_0; \theta_0)$. $\quad \square$

Next, we consider $d$ the distance on the parametric space $\Theta$, defined by

$$d(\theta, \theta') = \max\left\{|\alpha_j - \alpha_j'|, 1 \leq j \leq p, |\lambda - \lambda'|\right\}, \ \forall \ \theta, \theta' \in \Theta.$$

Recall that, from Proposition 2, we have

$$K(\theta) - K(\theta_0) = \mu_{\theta_0} \left(f(.; \theta) - f(.; \theta_0)\right)^2. \tag{36}$$

**Proof of Proposition 4**

Our aim is to prove that if there exists at least one $j \in \{1, 2, \ldots, p\}$ such that $\alpha_j^* \in \mathbb{R}\backslash\mathbb{Q}$, then for all (sufficiently small) $\varepsilon > 0$, we have

$$\inf_{\theta \in \Theta_\varepsilon} |K(\theta) - K(\theta_0)| > 0,$$

where $\Theta_\varepsilon = \{\theta : d(\theta, \theta_0) \geq \varepsilon\}$. We need distinguish three situations for the event $\Theta_\varepsilon$. Thus, we note that $\Theta_\varepsilon = \Gamma_1 \cup \Gamma_2 \cup \Gamma_3$, where

$$\Gamma_1 = \{\theta : \exists\, i \in \{1, \ldots, p\}, |\alpha_i - \alpha_i^*| \geq \varepsilon\}, \tag{37}$$

$$\Gamma_2 = \{\theta : \forall i \in \{1, \ldots, p\}, |\alpha_i - \alpha_i^*| < \varepsilon, |\lambda - \lambda^*| \geq \varepsilon, \langle\lambda\rangle \neq \langle\lambda^*\rangle\}, \tag{38}$$

and

$$\Gamma_3 = \{\theta : \forall i \in \{1, \ldots, p\}, |\alpha_i - \alpha_i^*| < \varepsilon, |\lambda - \lambda^*| \geq \varepsilon, \langle\lambda\rangle = \langle\lambda^*\rangle\}. \tag{39}$$

We are going to prove

$$\inf_{\theta \in \Gamma_i} |K(\theta) - K(\theta_0)| > 0, i = 1, 2, 3.$$

The idea of the proof is based on equation (36). Therefore, the aim is to find $x_0 \in E$ such that

$$|f(x_0; \theta) - f(x_0; \theta_0)| > 0, \text{ uniformly on } \Gamma_i, i = 1, 2, 3.$$

1. We consider the first case, $\theta \in \Gamma_1$. As $\Theta$ is compact implies that $|\lambda| \leq B$, where $B$ is a constant. Let $x = (0, \ldots, 0, x_i, 0, \ldots, 0) \in E$, we get

   $$|f(x; \theta) - f(x; \theta_0)| \geq \varepsilon |x_i| - 2B - 1.$$

   Therefore, there exists $y_0 > 0$ such that $\forall\, |x_i| \geq y_0$, we have $|f(x; \theta) - f(x; \theta_0)| \geq \frac{\varepsilon}{2}|x_i|$, uniformly on $\Gamma_1$. It follows, since the support of $\mu_{\theta_0}$ is not bounded,

   $$\inf_{\theta \in \Gamma_1} |K(\theta) - K(\theta_0)| > 0$$

2. For $\theta \in \Gamma_2$, let $0_E = (0, \ldots, 0) \in E$, we have

   $$|f(0_E; \theta) - f(0_E; \theta_0)| = |\langle\lambda\rangle - \langle\lambda^*\rangle| \geq 1.$$

   Because, if two real numbers do not have the same integer So that

   $$\inf_{\theta \in \Gamma_2} |K(\theta) - K(\theta_0)| > 0.$$

3. Finally, we consider the last case, $\theta \in \Gamma_3$. Here, we assume without loss of generality, $\alpha_1^* \in \mathbb{R}\backslash\mathbb{Q}$. Let $E' = \{x = (x_i) \in E : x_1 \in \mathbb{Z} \text{ and } x_j = 0, \ \forall\, j = 2, \ldots, p\}$.
   On $E'$, the actual **RINAR(p)** becomes a **RINAR(1)** model with $\alpha_1^* \in \mathbb{R}\backslash\mathbb{Q}$.
   Then, we retake the same proof of Proposition 4 from [12] for the similar case, remplacing in each step $x_0$ by a vector $y_0 = (x_0, 0, \ldots, 0)$. $\square$

**Proof of Theorem 1**

The conclusion $\hat{\theta}_n \to \theta_0$ almost surely results from Propositions 4 and 2 by standard arguments of the theory of M-estimators (see e.g. Van Der Vaart [17], Theorem 5.7). □

**Proof of Proposition 5**

Let $y = (x_1, \ldots, x_p)' \in E$. Then, from (17) (18) and (19), we have that for every $y \in E$ there exists $x \in \mathbb{Z}$ such that $\sum_{i=1}^{p} \alpha_i^* x_i = \nu_0 x$. Let $x \in \mathbb{Z}$. From (21), we have that $\nu_0 x = \dfrac{1}{\prod_{j=1}^{p} b_j}(d\ x)$ and again from (19) we get that there exists $y = (x_1, \ldots, x_p)' \in E$ such that $d\ x = A_1 x_1 + \ldots + A_p x_p$. It is follows that

$$\nu_0 x = \frac{1}{\prod_{j=1}^{p} b_j}(A_1 x_1 + \ldots + A_p x_p).$$

Finally, from (17), we get that for every $x \in \mathbb{Z}$ there s $y = (x_1, \ldots, x_p)' \in E$ such that $\nu_0 x = \sum_{i=1}^{p} \alpha_i^* x_i$. □

**Proof of Proposition 6**

Note that $K(\theta) = K(\theta_0)$ means, from Proposition 2, $f(.;\theta) = f(.;\theta_0) \ \mu_{\theta_0} - a.s$
i.e. $\forall\ y = (x_1, \ldots, x_j, \ldots, x_p)' \in E$, we have

$$\langle \sum_{i=1}^{p} \alpha_i x_i + \lambda \rangle = \langle \sum_{i=1}^{p} \alpha_i^* x_i + \lambda^* \rangle.$$

Let $y_i = (0, \ldots, 0, x_i, 0, \ldots, 0)' \in E$ for $i = 1, \ldots, p$. Letting $|x_i| \to \infty$ for each $y_i$, we find

$$\alpha_i = \alpha_i^* \ \forall\ i = 1, \ldots, p.$$

So, from Proposition 5, we get

$$\forall\ y \in E,\ f(y;\theta) = f(y;\theta_0) \iff \forall\ x \in \mathbb{Z},\ \langle \nu_0 x + \lambda \rangle = \langle \nu_0 x + \lambda^* \rangle.$$

It is follows, from Proposition 5 of [12], that

$$\forall\ x \in \mathbb{Z},\ \langle \nu_0 x + \lambda \rangle = \langle \nu_0 x + \lambda^* \rangle \iff \lambda \in I_0. \ \square$$

Let $x^0 = (x_1^0, \ldots, x_p^0)' \in E$. By Bézout theorem, we know that there exits $x_0 \in \mathbb{Z}$ such that $\sum_{i=1}^{p} \alpha_i^* x_i^0 = \nu_0 x_0$. The following lemma will be useful for the proof of Proposition 7.

**Lemma 3.** *The function* $\alpha = (\alpha_1, \ldots, \alpha_p) \longrightarrow \left\{ \sum_{j=1}^{p} \alpha_j x_j^0 \right\}$, *defined on* $\mathbb{R}^p$, *is continuous at* $\alpha_0 = (\alpha_1^*, \ldots, \alpha_p^*)$ *if* $x_0 \neq mb$, *with* $m \in \mathbb{N}$ *and* $b$ *is denumerator of irreductible fraction form of* $\nu_0$.

**Proof of Lemma 3**

Let $x^0 = (x_1^0, \ldots, x_p^0) \in E$ fixed. We know that the function $h : \alpha \longrightarrow \{\alpha \cdot x^0\} = \left\{\sum_{j=1}^p \alpha_j x_j^0\right\}$, defined on $\mathbb{R}^p$ and with values in $[0, 1[$, is discontinuous at $\alpha_0$, if $\sum_{j=1}^p \alpha_j^* x_j^0 \in \mathbb{Z}$. From Bézout theorem, there exists $x_0 \in \mathbb{Z}$ fixed, such that

$$\sum_{i=1}^p \alpha_i^* x_i^0 = \nu_0 x_0 = \frac{a}{b} x_0.$$

So, if $x_0 \neq mb$, we get $\nu_0 x_0 \notin \mathbb{Z}$, it follows $h$ is continuous at $\alpha_0$. $\square$

Recall that, $d$ the distance on the parametric space $\Theta$, is defined by

$$d(\theta, \theta') = \max\left\{|\alpha_j - \alpha_j'|, 1 \leq j \leq p, |\lambda - \lambda'|\right\}, \ \forall \ \theta, \theta' \in \Theta.$$

**Proof of Proposition 7**

We have, $\nu_0 = \frac{a}{b}$ where $a \in \mathbb{Z}$ and $b \in \mathbb{Z}_{>0}$ ($b$ is even). $a$ and $b$ are comprime.

Thus, there exists $k_0 \in \{0, 1, \ldots, q-1\}$ such that

$$\{\lambda^*\} \in i_0 = \left[\frac{k_0}{b}, \frac{k_0 + 1}{b}\right[.$$

We recall,

$$I_0 = \{\lambda : \langle\lambda\rangle = \langle\lambda^*\rangle, \{\lambda\} \in i_0\} \text{ and } E_0 = \{\theta \in \Theta : \alpha_j = \alpha_j^*, \ \forall \ j = 1, \ldots, p \text{ and } \lambda \in I_0\}.$$

Our aim is to prove that for all (sufficiently small) $\varepsilon > 0$, we have

$$\inf_{\theta \in \Theta_\varepsilon^0} |K(\theta) - K(\theta_0)| > 0,$$

where $\Theta_\varepsilon^0 = \{\theta : d(\theta, E_0) \geq \varepsilon\}$. We need distinguish three situations for the event $\Theta_\varepsilon^0$. We note that $\Theta_\varepsilon^0 = \Gamma_1 \cup \Gamma_2 \cup \Gamma_3$, where

$$\Gamma_1 = \{\theta : \exists \ i \in \{1, \ldots, p\}, |\alpha_i - \alpha_i^*| \geq \varepsilon\}, \tag{40}$$

$$\Gamma_2 = \{\theta : \forall i \in \{1, \ldots, p\}, |\alpha_i - \alpha_i^*| < \varepsilon, d(\lambda, I_0) \geq \varepsilon, \langle\lambda\rangle \neq \langle\lambda^*\rangle\}, \tag{41}$$

and

$$\Gamma_3 = \{\theta : \forall i \in \{1, \ldots, p\}, |\alpha_i - \alpha_i^*| < \varepsilon, d(\lambda, I_0) \geq \varepsilon, \langle\lambda\rangle = \langle\lambda^*\rangle\}. \tag{42}$$

We are going to prove

$$\inf_{\theta \in \Gamma_i} |K(\theta) - K(\theta_0)| > 0, i = 1, 2, 3.$$

From Proposition 2, we know that

$$K(\theta) - K(\theta_0) = \mu_{\theta_0} \left(f(.; \theta) - f(.; \theta_0)\right)^2.$$

So, the aim is to find $x_0 \in E$ such that

$$|f(x_0; \theta) - f(x_0; \theta_*)| > 0, \ \forall \ \theta_* \in \Gamma_i, i = 1, 2, 3.$$

1. For $\theta \in \Gamma_1 \cup \Gamma_2$, then for the same arguments of the equivalents subsets of the proof of Proposition 4 we have that
$$\inf_{\theta \in \Gamma_1 \cup \Gamma_2} |K(\theta) - K(\theta_0)| > 0.$$

2. For $\theta \in \Gamma_3$, without loss of generality we assume that $\nu_0 > 0$, $\lambda^* \geq 0$. Note that
$$d(\theta, E_0) = d(\lambda, I_0) = \inf_{\lambda_* \in I_0} |\lambda - \lambda_*|.$$

We distinguish four cases depending on the position of the fractional part of $\lambda^*$ and $\lambda$:

(a) Case A : $i_0 \subset \left[0, \frac{1}{2}\right[$ and $\lambda \geq I_0$.

(b) Case B : $i_0 \subset \left[0, \frac{1}{2}\right[$ and $\lambda \leq I_0$.

(c) Case C : $i_0 \subset \left[\frac{1}{2}, 1\right[$ and $\lambda \geq I_0$.

(d) Case D : $i_0 \subset \left[\frac{1}{2}, 1\right[$ and $\lambda \leq I_0$.

- Case A : $i_0 \subset \left[0, \frac{1}{2}\right[$ ; $\lambda \geq I_0$. For this case, it is simple to verified that
$$b \geq 4, \quad \{\lambda\} \in \left[0, \frac{1}{2}\right[, k_0 \in \left\{0, 1, \ldots, \frac{b}{2} - 2\right\} \text{ and } \{\lambda\} \geq \frac{k_0 + 1}{b}.$$

Therefore,
$$d(\theta, E_0) = \inf_{\lambda_* \in i_1} \{\lambda\} - \{\lambda_*\} \geq \varepsilon. \tag{43}$$

There exists $n \in \{k_0 + 1, \ldots, \frac{b}{2} - 1\}$ such that $\{\lambda\} \in \left[\frac{n}{b}, \frac{n+1}{b}\right[$. Let $x_0 = \frac{b}{2} - n$ fixed. We know that, from Bézout theorem, for $x_0$ used here there exits $x^0 = (x_1^0, \ldots, x_p^0) \in E$ such that
$$\sum_{j=1}^p \alpha_j^* x_j^0 = \nu_0 x_0.$$

For all $\{\lambda_*\} \in i_0$ we have
$$\left\{\sum_{j=1}^p \alpha_j^* x_j^0\right\} + \{\lambda_*\} = \left\{\frac{1}{b} x_0\right\} + \{\lambda_*\} < \frac{1}{2}, \tag{44}$$

and from (43) we get
$$\left\{\sum_{j=1}^p \alpha_j^* x_j^0\right\} + \{\lambda\} = \left\{\frac{1}{b} x_0\right\} + \{\lambda\} \geq \frac{1}{2} + \varepsilon. \tag{45}$$

Also, for $x_0 = \frac{b}{2} - n$, from Lemma 3, the function $\alpha = (\alpha_1, \ldots, \alpha_p) \longrightarrow \left\{\sum_{j=1}^p \alpha_j x_j^0\right\}$ is continuous in $\alpha_0 = (\alpha_1^*, \ldots, \alpha_p^*)$. There exists $\eta = \eta(\varepsilon, \nu_0, \lambda^*) \leq \varepsilon$ such that $\forall j = 1, \ldots, p \ |\alpha_j - \alpha_j^*| \leq \eta$ it arise
$$\begin{cases} [\sum_{j=1}^p \alpha_j x_j^0] = [\sum_{j=1}^p \alpha_j^* x_j^0]. \\ |\left\{\sum_{j=1}^p \alpha_j x_j^0\right\} - \left\{\sum_{j=1}^p \alpha_j^* x_j^0\right\}| \leq \varepsilon. \end{cases} \tag{46}$$

From (45) and (46) we get

$$\left\{\sum_{j=1}^{p}\alpha_j x_j^0\right\} + \{\lambda\} = \left(\left\{\sum_{j=1}^{p}\alpha_j x_j^0\right\} - \left\{\sum_{j=1}^{p}\alpha_j^* x_j^0\right\}\right) + \left(\left\{\sum_{j=1}^{p}\alpha_j^* x_j^0\right\} + \{\lambda\}\right)$$
$$\geq -\varepsilon + \left(\frac{1}{2}+\varepsilon\right) = \frac{1}{2}.$$

Then

$$\begin{aligned}
f(x^0;\theta) - f(x^0;\theta_0) &= \langle\sum_{j=1}^{p}\alpha_j x_j^0 + \lambda\rangle - \langle\sum_{j=1}^{p}\alpha_j^* x_j^0 + \lambda_*\rangle \\
&= \left[\sum_{j=1}^{p}\alpha_j x_j^0\right] - \left[\sum_{j=1}^{p}\alpha_j^* x_j^0\right] + \langle\lambda\rangle - \langle\lambda_*\rangle \\
&\quad + \langle\left\{\sum_{j=1}^{p}\alpha_j x_j^0\right\} + \{\lambda\}\rangle - \langle\left\{\sum_{j=1}^{p}\alpha_j^* x_j^0\right\} + \{\lambda_*\}\rangle \\
&= 1
\end{aligned}$$

- The other cases :
    - Case B : $i_0 \subset \left[0, \frac{1}{2}\right[$ and $\lambda \leq I_0$.
    - Case C : $i_0 \subset \left[\frac{1}{2}, 1\right[$ and $\lambda \geq I_0$.
    - Case D : $i_0 \subset \left[\frac{1}{2}, 1\right[$ and $\lambda \leq I_0$.

    can be treated in a similar way, with a suitable choice of $x_0$, and $\alpha = (\alpha_1, \ldots, \alpha_p)$ sufficiently close to $\alpha_0 = (\alpha_1^*, \ldots, \alpha_p^*)$. Then, we deduced that there exists $\eta = \eta(\varepsilon, \nu_0, \lambda^*) \leq \varepsilon$ and a constant $e_0 > 0$, uniformly on $\Gamma_3' = \{\theta : \theta \in \Gamma_3, \forall\, j \in \{1, \ldots, p\}\ |\alpha_j - \alpha_j^*| \leq \eta\}$, we have $|K(\theta) - K(\theta_0)| \geq e_0 > 0$.
    For $\theta \in \Gamma_3'' = \Gamma_3 \setminus \Gamma_3' = \{\theta : \theta \in \Gamma_3, \forall\, j \in \{1, \ldots, p\}\ |\alpha_j - \alpha_j^*| > \eta\}$ the same argument for $\Gamma_1$ can be applied here and then there exists $d_0 > 0$ such that

    $$\inf_{\theta \in \Gamma_3''} |K(\theta) - K(\theta_0)| \geq d_0 > 0. \quad \square$$

**Proof of Theorem 2**

The conclusion $\hat{\theta}_n \to E_0$, almost surely, results from Propositions 2 and 7 by the same arguments used in the proof of Theorem 1. $\square$

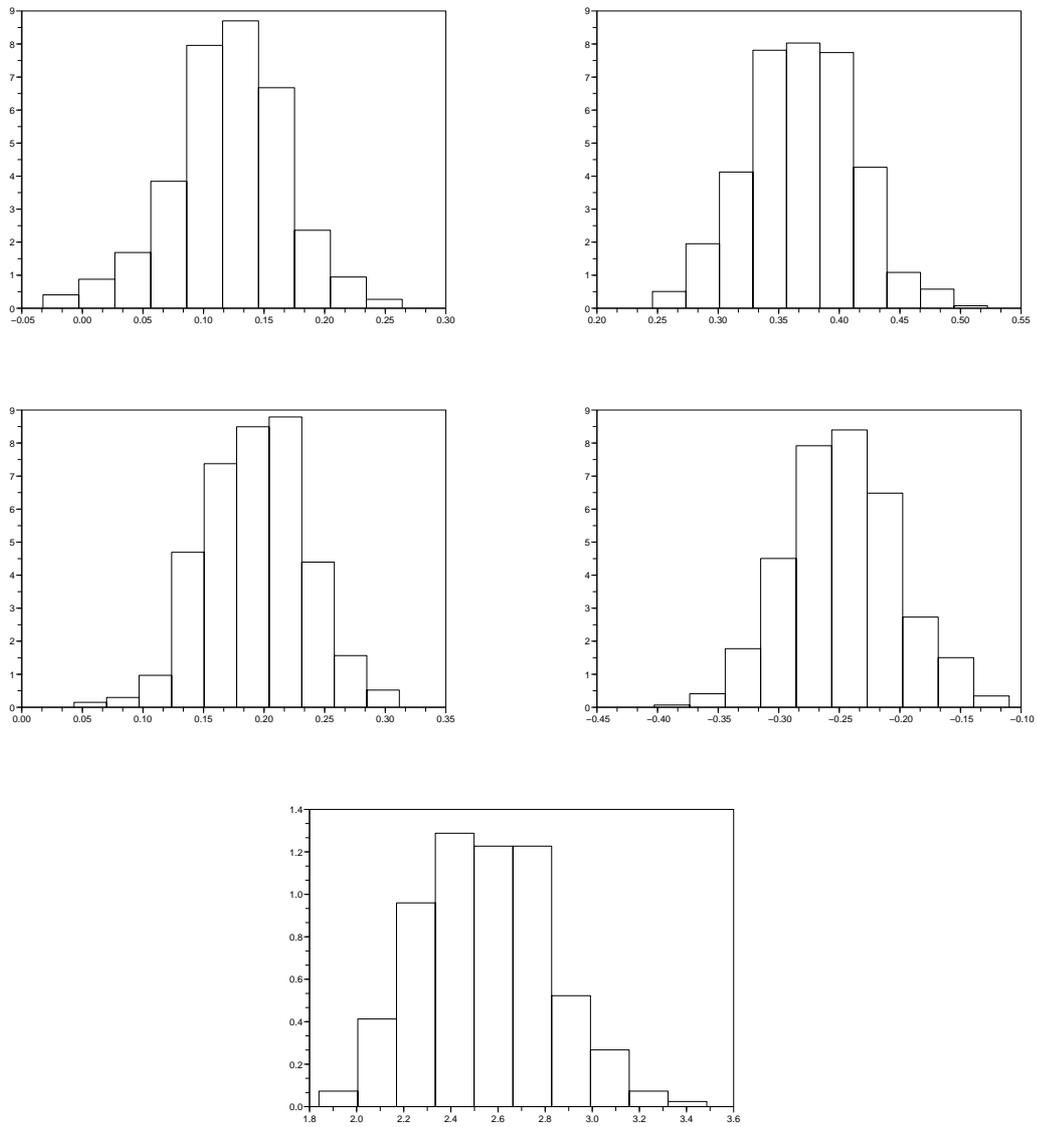

Figure 2: Histograms of the parameters estimates $\hat{\alpha}_j, j = 1, \ldots, 4$ and $\hat{\lambda}$ (left to right, top to bottom). 500 replications of time series of length 500. Actual values of the parameters are $\theta_0 = \left(\frac{3}{25}, \frac{3}{8}, \frac{1}{5}, \frac{-1}{4}, \frac{5}{2}\right)$.

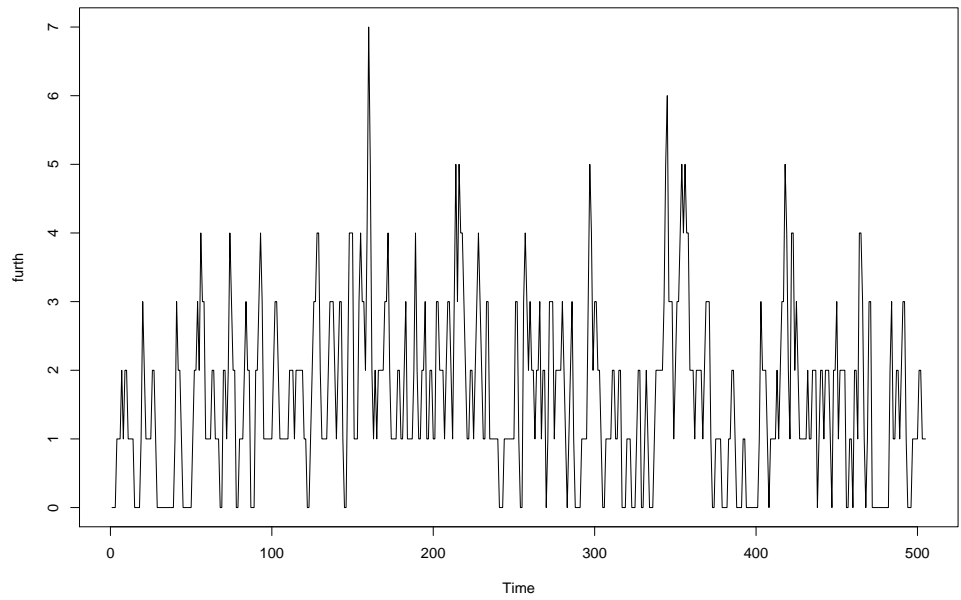

Figure 3: Plot of the Fürth data.

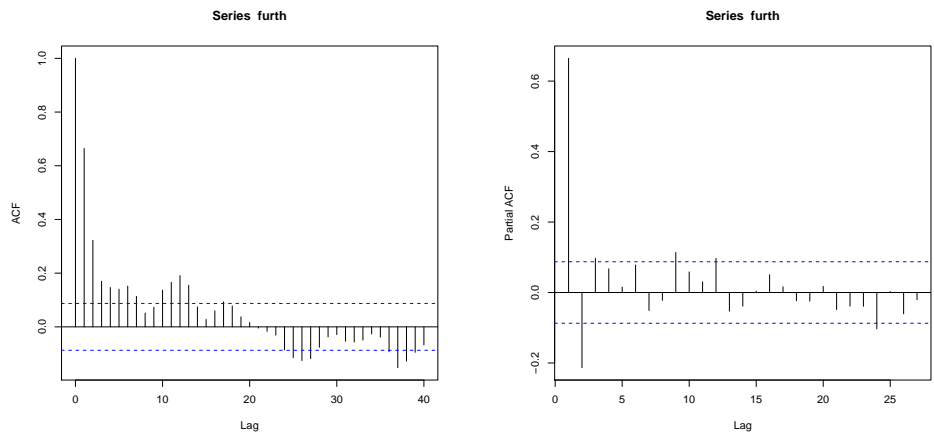

Figure 4: Sample autocorrelation and partial autocorrrelation functions of Fürth data.

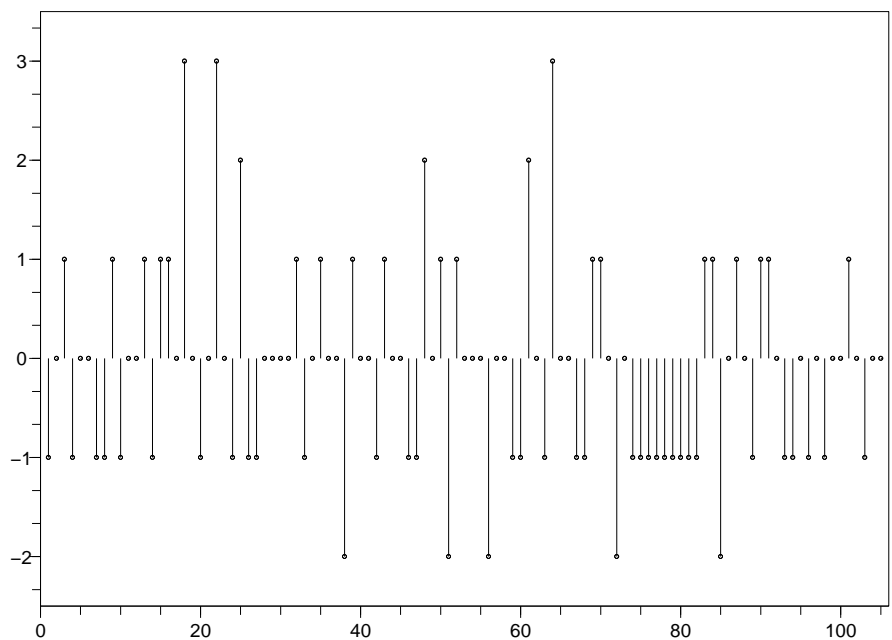

Figure 5: The forecast errors $\hat{\varepsilon}_{T+1} = X_{T+1} - \hat{X}_{T+1}$, $400 \leq T \leq 504$, of the Fürth data, from a **RINAR(2)** fit.